\documentstyle[sprocl]{article}

\input{psfig.sty}
\newcommand{\mincir}{\raise -2.truept\hbox{\rlap{\hbox{$\sim$}}\raise5.truept
\hbox{$<$}\ }}
\newcommand{\magcir}{\raise -2.truept\hbox{\rlap{\hbox{$\sim$}}\raise5.truept
\hbox{$>$}\ }}
\newcommand{\minmag}{\raise-2.truept\hbox{\rlap{\hbox{$<$}}\raise 6.truept\hbox
{$>$}\ }}

\def\be{\begin{equation}}
\def\ee{\end{equation}}
\def\bea{\begin{eqnarray}}
\def\eea{\end{eqnarray}}

\begin{document}

\title{ DARK MATTER AND THE SCALE OF SUSY BREAKING}

\author{ S. BORGANI$^{1,2}$ and A. MASIERO$^{1,2,3}$ }

\address{$(1)$ INFN, Sezione di Perugia, c/o Dipartimento di Fisica
dell'Universit\`a, \\via A. Pascoli, I-06100 Perugia, Italy} 
\address{$(2)$ SISSA, International School for Advanced Studies, \\
via Beirut 2, I-34015 Trieste, Italy} 
\address{$(3)$ Dipartimento di Fisica dell'Universit\`a di Perugia, \\
via A. Pascoli, I-06100 Perugia, Italy}

\maketitle 

\abstracts{ We consider the prospects for dark matter (DM) in two
classes of supersymmetric (SUSY) models which are characterized by the
different mechanism of SUSY breaking, namely the more common
supergravity models with very large scale of SUSY breaking and the
recent schemes where SUSY is broken at a relatively low scale and
gravitinos are likely to be the lightest SUSY particle. We point out
that the former scheme is in general associated with the cold dark
matter (CDM) scenario, while the latter predicts a warm dark matter
(WDM) dominated Universe. }

\section{Introduction} 

We have strong indications that ordinary matter
(baryons) is insufficient to provide the large amount of non-shining
matter (dark matter DM) which has been observationally proved to exist
in galactic halos and at the level of clusters of galaxies
\cite{kolb}. In a sense, this might constitute the ``largest"
indication of new physics beyond the standard model (SM) for
electroweak interactions. This statement holds true even after the
recent stunning developments in the search for non-shining baryonic
objects. In September 1993 the discovery of massive dark objects
(``machos") was announced. After three years of intensive analysis it
is now clear that in any case machos cannot account for the whole dark
matter of the galactic halos \cite{roulet}.

It was widely expected that some amount of non--shining baryonic
matter could exist given that the contribution of luminous ba\-ryons
to the energy density of the Universe $\Omega_0$ = $\rho/\rho_{cr}$
($\rho_{cr}= 3H^2_0 /8 \pi G$ where G is the gravitational constant
and $H_0$ the Hubble constant) is less than $1\%$, while from
nucleosynthesis we infer $\Omega_{baryon} = \rho_{baryon} / \rho_{cr}
= (0.06\- \pm 0.02) h_{50}^{-2}$, where $h_{50} = H_0 /(50 Km s^{-1}
Mpc^{-1})$.  On the other hand, we have direct indications\cite{Dekel}
that $\Omega_0$ should be at least $20\%$ which means that baryons can
represent no more than half of the entire energy density of the
Universe \cite{kolb}.

We could make these considerations about the insufficiency of the SM
to obtain a large enough $\Omega_0$ more dramatic if we accept the
theoretical input that the Universe underwent some inflationary era
which produced $\Omega_0$ extremely close to unity. In that case, at
least $90\%$ of the whole energy density of the Universe should be
provided by some new physics beyond the SM.

Before discussing possible particle physics candidates, it should be
kept in mind that DM is not only called for to provide a major
contribution to $\Omega_0$, but also it has to provide a suitable
gravitational driving force for the primordial density fluctuations to
evolve into the large-scale structures (galaxies, clusters and
superclusters of galaxies) that we observe today \cite{kolb}. Here we
encounter the major difficulties when dealing with the two
``traditional" sources of DM: cold (CDM) and hot (HDM) DM.

Light neutrinos in the eV range are the most typical example of HDM, being
their decoupling temperature of O(1 MeV). On the other hand, the lightest
supersymmetric particle (LSP) in the tens of GeV range is a typical CDM
candidate. Taking the LSP to be the lightest neutralino, one obtains that
when it decouples it is already non-relativistic, being its decoupling
temperature typically one order of magnitude below its mass. 

Both HDM and CDM have some difficulty to correctly reproduce
observations related to the distribution of cosmic structures
(galaxies and galaxy clusters) at different scales and at different
redshifts. The conflict is more violent in the case of pure HDM.
Neutrinos of few eV's erase by free--streaming density fluctuations on
small (galactic) scales, thus producing a wrong galaxy clustering
pattern, as well as a too late galaxy formation.  The opposite problem
arises with pure CDM: we obtain too much power in the spectrum at small
mass scales.

A general feature is that some amount of CDM should be present in any case.
A possibility which has been envisaged is that after all the whole
$\Omega_0$ could be much smaller than one, say $20\%$ or so and then entirely
due to CDM. However, if one keeps on demanding the presence of an
inflationary epoch, then, according to the standard picture, 
it seems unnatural to have $\Omega_0$ so different from unity.

Another possibility is that CDM provides its $20\%$ to $\Omega_0$, while 
flatness is provided by a non vanishing cosmological
constant. Needless to say, not having even a satisfactory reason for a
vanishing cosmological constant, it seems harder to naturally obtain some
specific non-vanishing value for it.

Finally, a further possibility is given by the so-called mixed dark
matter (MDM) scenario\cite{shafi}, where a wise cocktail of HDM and
CDM is present. An obvious realization of a MDM scheme is a variant of
the minimal supersymmetric standard model (MSSM)\cite{nilles} where
neutrinos get a mass of few eV's. In that case the lightest neutralino
(which is taken to be the LSP) plays the role of CDM and the light
neutrino(s) that of HDM. With an appropriate choice of the parameters
it is possible to obtain contributions to $\Omega_0$ from the CDM and
HDM in the desired range.

In the first part of this talk we will briefly review the 
``standard" situation which occurs in supergravity models where local SUSY
is broken at a very high scale and gravity constitutes the messenger of
SUSY breaking to the observable sector. In this frame the LSP is likely
to be the lightest neutralino with typical mass in the tens of GeV. In the
second part we will move to newer
schemes where the breaking of SUSY occurs at scale not far from the
electroweak scale and gauge forces instead of gravity are invoked as
messengers of the SUSY breaking to the observable sector. This frame of
SUSY breaking entails major differences on the DM problem. Indeed,
gravitinos turn out to be quite light in this context and they are
likely to constitute the new LSP. We will discuss the implications of
these gauge-mediated SUSY breaking models for DM schemes.

\section{High-Scale Susy Breaking and DM}
In N=1 supergravity models \cite{cremmer}, where a discrete symmetry, matter
R--parity,  discriminates
between ordinary and SUSY particles, the lightest SUSY particle
(LSP) is absolutely stable. For several reasons the lightest
neutralino is the fa\-vourite candidate to be the LSP fulfilling
the role of CDM \cite{jungman}.

The neutralinos are the eigenvectors of the mass matrix of the four 
neutral fermions partners of the $W_3, B, H^0_1$ and $H^0_2$. There
are four parameters entering this matrix: $M_1, M_2, \mu$ and $tg\beta$.
The first two parameters denote the coefficients of the SUSY breaking 
mass terms $\tilde B \tilde B$ and $\tilde W_3 \tilde W_3$ respectively.
$\mu$ is the coupling of the $H_1 - H_2$ term the superpotential.
Finally  $tg \beta$ denotes the ratio of the VEV's of the $H_2$ and
$H_1$ scalar fields. 

In general $M_1$ and $M_2$ are two independent
parameters, but if one assumes that a grand unification scale takes place,
then at the grand unification $M_1 = M_2 = M_3$, where $M_3$ is the
gluino mass at that scale. Then at $M_W$ one obtains:

\bea
M_1 &=& {5 \over 3} tg^2 \theta_w M_2 \simeq {M_2 \over 2},\nonumber \\
M_2 &=& {g^2_2 \over g^2_3} m_{\tilde g} \simeq m_{\tilde g} /3,
\label{m12}
\eea
where $g_2$ and $g_3$ are the SU(2) and SU(3) gauge coupling constants, 
respectively.

The relation (\ref{m12}) between $M_1$ and $M_2$ reduces to three the
number of independent parameters which determine the lightest
neutralino composition and mass: $tg \beta, \mu$ and $M_2$. Hence, for
fixed values of $tg \beta$ one can study the neutralino spectrum in
the ($\mu, M_2$) plane. The major experimental inputs to exclude
regions in this plane are the request that the lightest chargino be
heavier than $M_Z /2$ (actually, this limit has been now pushed up to
$\sim 80$ GeV with the advent of LEP200) and the limits on the
invisible width of the Z hence limiting the possible decays $Z
\rightarrow \chi \chi,\-\chi \chi'$.

Moreover if the GUT assumption is made, then the relation (\ref{m12}) 
between $M_2$ and $m_{\tilde g}$ implies a severe bound on $M_2$ from 
the experimental lower bound on  $m_{\tilde g}$ from Tevatron
(roughly $m_{\tilde g}$
$> 150 GeV$, hence implying $M_2 > 50\-GeV)$. The theoretical demand
that the electroweak symmetry be broken radiatively , i.e. due to
the renormalization effects on the Higgs masses when going from the 
superlarge scale of supergravity breaking down to $M_W$, further 
constrains the available ($\mu, M_2$) region.

The first important outcome of this analysis is that the lightest 
neutralino mass exhibits a lower bound of roughly 20 GeV .
The prospects for an improvement of this lower limit at LEP 200
crucially depends on the composition of $\chi$. If $\chi$ is 
mainly 
a gaugino, then it is difficult to go beyond 40 GeV for such a
lower bound, while with a $\chi$ mainly higgsino the lower bound
can jump up to $m_\chi > M_W$ at LEP 200.

Let us focus now on the role played by $\chi$ as a source of CDM. 
$\chi$ is kept in thermal equilibrium through its electroweak
interactions not only for $T > m_\chi$, but even when T is
below $m_\chi$. However for $T < m_\chi$ the number of $\chi's$
rapidly decreases because of the appearance of the typical 
Boltzmann suppression factor exp ($- m_\chi /T$). When T is
roughly $m_\chi /20$ the number of $\chi$ diminished so much
that they do not interact any longer, i.e. they decouple.
Hence the contribution to $\Omega_{CDM}$ of $\chi$ is determined by two 
parameters: $m_\chi$ and the temperature at which $\chi$ decouples
$(T_D)$. $T_D$ fixes the number of $\chi's$ which survive. As
for the determination of $T_D$ itself, one has to compute the
$\chi$ annihilation rate and compare it with the cosmic expansion rate
\cite{ellis}.

Several annihilation channels are possible with the exchange
of different SUSY or ordinary particles, $\tilde f$, H, Z,
etc. . Obviously the relative importance of the channels 
depends on the composition of $\chi$. For instance, if 
 $\chi$ is a pure gaugino, then the $\tilde f$ exchange represents
the dominant annihilation mode. 

Quantitatively \cite{bottino}, it turns out that if $\chi$ results from a
large  mixing of the gaugino ($\tilde W_3$ and $\tilde B$) and higgsino
($\tilde H^0_1$ and $\tilde H^0_2$) components, then the
annihilation is too efficient to allow the surviving $\chi$ to provide 
$\Omega_0$ large enough. Typically in this case $\Omega_0 < 10^{-2}$
and hence $\chi$ is not a good CDM candidate. On the contrary, 
if $\chi$ is either almost a pure higgsino or a pure gaugino
then it can give a conspicuous contribution to $\Omega_0$

In the case $\chi$ is mainly a gaugino (say at least at the $90 \%$
level) what is decisive to establish the annihilation rate is the mass
of $\tilde f$.  LEP 200 will be able, hopefully, to test slepton
masses up to $M_W$.  If sfermions are light, the $\chi$ annihilation
rate is fast and the $\Omega_\chi$ is negligible.  On the other hand,
if $\tilde f$ (and hence $\tilde l$, in particular) is heavier than
150 GeV, the annihilation rate of $\chi$ is sufficiently suppressed so
that $\Omega_\chi$ can be in the right ball-park for $\Omega_{CDM}$. In
fact if all the $\tilde f's$ are heavy, say above 500 GeV and for
$m_\chi \ll m_{\tilde f}$, then the suppression of the annihilation
rate can become even too efficient yielding $\Omega_\chi$ unacceptably
large.

Let us briefly discuss the case of $\chi$ being mainly a higgsino. 
If the
lightest neutralino is to be predominantly a combination of
$\tilde H_1^0$ and $\tilde H^0_2$ it means that $M_1$ and $M_2$ have
to be much larger than $\mu$. Invoking the relation (\ref{m12}) one
concludes that in this case we expect heavy gluinos, typically
in the TeV range. As for the number of surviving $\chi's$ in 
this case, what is crucial is whether $m_\chi$ is larger or
smaller than $M_W$.
Indeed, for $m_\chi > M_W$ the annihilation channels $\chi \chi$
$\rightarrow W W, ZZ, t \bar t$ reduce $\Omega_\chi$ too much. 
If $m_\chi < M_W$ then acceptable contributions of $\chi$ to
$\Omega_{CDM}$ are obtainable in rather wide areas of the
($\mu - M_2$) parameter space. Once again we emphasize that the case
$\chi$ being a pure higgsino is of particular relevance for LEP 200 given 
that in this case $\chi$ masses up to $M_W$ can be explored.

In the minimal SUSY standard model there are five new parameters 
in addition to those already present in the non--SUSY case.
Imposing the electroweak radiative breaking further reduces
this number to four. Finally, in simple supergravity realizations
the soft parameters A and B are related. Hence we end up with 
only three new independent parameters. One can use the constraint
that the relic $\chi$ abundance provides a correct $\Omega_{CDM}$
to restrict the allowed area in this 3--dimensional space.
Or, at least, one can eliminate points of this space which would
lead to $\Omega_\chi >1$, hence overclosing the Universe. For
$\chi$ masses up to 150 GeV it is possible to find sizable regions
in the SUSY parameter space where $\Omega_\chi$ acquires interesting
values for the DM problem. A detailed discussion on this point is beyond
the scope of this talk. The interested reader can find a thorough analysis
in the review of Ref.~\cite{jungman} and the original papers therein quoted.

\section{Light Gravitinos as Dark Matter}
An alternative scenario to that we discussed in the previous Section
is based on the possibility that SUSY is broken in a ``secluded"
sector at a much lower scale with gauge instead of gravitational
forces responsible for conveying the breaking of SUSY to the
observable sector. This scenario had already been critically
considered in the old days of the early constructions of SUSY models
and has raised a renewed interest recently with the proposal by
Refs.~\cite{dineetal,dine-old,dvalietal}, where some guidelines for the
realization of low-energy SUSY breaking are provided.  In these
schemes, the gravitino mass ($m_{3/2}$) loses its role of fixing the
typical size of soft breaking terms and we expect it to be much
smaller than what we have in models with a hidden sector.  Indeed,
given the well-known relation \cite{nilles} between $m_{3/2}$ and the
scale of SUSY breaking $\sqrt{F}$, i.e.\ $m_{3/2}=O(F/M)$, where $M$
is the reduced Planck scale, we expect $m_{3/2}$ in the keV range for
a scale $\sqrt{F}$ of $O(10^6$ GeV) that has been proposed in models
with low-energy SUSY breaking in a visible sector.

Models with light gravitinos have recently attracted much
phenomenological attention \cite{ambrosanio} also in relation with the
``famous" anomalous CDF $\gamma \gamma e^+ e^-$ event.  Needless to
say, one should be very cautious in drawing any implication from just
a single event. However, with this reservation in mind, we can surely
state that this event finds an attractive interpretation in the decay
of neutralinos having light gravitinos in the final products.

In the following we briefly report some implications of SUSY models
with a light gravitino (in the keV range) in relation with the dark
matter (DM) problem. We anticipate that a gravitino of that mass
behaves as a warm dark matter (WDM) particle, that is, a particle
whose free streaming scale involves a mass comparable to that of a
galaxy, $\sim 10^{11-12}M_\odot$.

Suppose that the gravitinos were once in thermal equilibrium and were
frozen out at the temperature $T_f$ during the cosmic expansion. 
It can be shown that the density parameter $\Omega_0$ 
contributed by relic thermal gravitinos is
\be
\Omega_0 h^2 =1.17 \left( {m_{3/2}\over 1\mbox{keV}} \right) 
               \left( {g_{*}(T_f)\over 100} \right)^{-1} ,
\ee
where $g_*(T_f)$ represents the effective massless degrees of freedom
at the temperature $T_f$.  Therefore, a gravitino in the
above-mentioned keV range provides a significant portion of the mass
density of the present Universe.

As for the redshift at which gravitinos becomes non relativistic, it
corresponds to the epoch at which their temperature becomes
$m_{3/2}/3$. That is,
\bea
Z_{nr} & \simeq &  \left( \frac{g_*(T_f)}{g_{*S}(T_0)} \right)^{1/3}
     \frac{m_{3/2}/3}{T_0} \\ \nonumber
       & = & 4.14 \times 10^6 \times  
     \left( \frac{g_*(T_f)}{100} \right)^{1/3}
     \left( \frac{m_{3/2}}{1\mbox{keV}} \right)\,,
\eea
where $T_0=2.726\,K$ is the temperature of the CMB at the present time.
Once $Z_{nr}$ is known, one can estimate  the free streaming length until the
epoch of the matter-radiation equality, $\lambda_{FS}$, which represents
a quantity of crucial relevance for the formation of large--scale cosmic
structures. If $v(t)$ is the typical velocity of a DM particle at the
time $t$, then
\bea
\lambda_{FS}& \equiv & \int_0^{t_{eq}} {v(t)\over a(t)} dt \\ \nonumber
                 & = & 2 t_0 \times {Z_{eq}^{1/2}\over Z_{nr}}
                         [1+\ln (Z_{nr}/Z_{eq})] \\ \nonumber
                 & = & 6.08 \times 10^5 \times Z_{nr}^{-1} \mbox{Mpc}
                  [1+\ln(Z_{nr}/2.32 h^2 \times 10^4)]\,,
\label{eq:free} 
\eea 
where $Z_{eq}$ is the redshift of
matter--radiation equality.  Accordingly, the free-streaming length
for the thermal gravitinos is about 1Mpc (for $Z_{nr}\sim 4 \times
10^6$), which in turn corresponds to $\sim10^{12}M_\odot$, if it is
required to provide a density parameter close to unity. This
explicitly shows that light gravitinos are actually WDM candidates. We
also note that, taking $h=0.5$, the requirement of not overclosing the
Universe turns into $m_{3/2}\mincir 200\,eV$. Quite interestingly,
this constraint on $m_{3/2}$ is of the same order as the upper limit
of $\sim 250\,eV$, based on the light gravitino interpretation of the
mentioned CDF event\cite{ambrosanio}.  Moreover, if we rely on the
lower bound, $m_{\tilde g}\ge 30$ eV, based on observations about the
energy loss from the supernova SN1987A\cite{Grifols}, then we conclude
that $\Omega_0\magcir 0.15$ should be provided by light gravitinos.

However, critical density models with pure WDM are known to suffer for
serious troubles\cite{colo}. Indeed, a WDM scenario
behaves much like CDM on scales above $\lambda_{FS}$. Therefore,
we expect in the light gravitino scenario that the level of
cosmological density fluctuations on the scale of galaxy clusters
($\sim 10\,h^{-1}$Mpc) to be almost the same as in CDM. As a
consequence, the resulting number density of galaxy clusters is
predicted to be much larger than what observed\cite{clth}).

\begin{figure}
{\centerline{
\psfig{figure=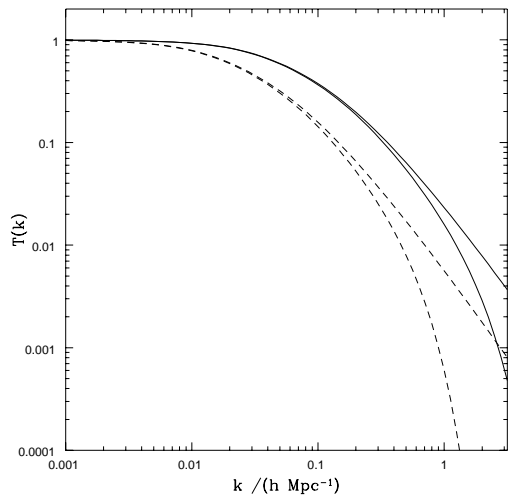,height=7.truecm}}}
\caption{Transfer functions for WDM (heavier lines) and CDM (lighter
lines) by using the fitting function by Bardeen et al.
Solid and dashed curves are for $(\Omega_0,h)=(1,0.5)$ and $(0.3,0.7)$
respectively. The free--streaming scale for WDM is chosen to be 
$\lambda_{FS}=0.5\,\Omega_0 h^{-1}$Mpc.}
\label{fig:tk}
\end{figure}

In order to overcome these problems one should follow in
principle the same patterns as for improving the standard CDM
scenario: {\it (a)} tilting the primordial spectrum to $P(k)\propto
k^n$ with $n<1$, also assuming a large ($\sim 20\%$) baryon fraction
\cite{vianaetal}; {\it (b)} decreasing the density parameter to
$\Omega_0<1$, with the possible introduction of a cosmological
constant term to make the spatial curvature negligible, as predicted by
standard inflationary models; {\it (c)} add a second DM component,
made by particles with a much larger $\lambda_{FS}$,so as to suppress
fluctuations on the cluster scales.

As for the scenario {\it (b)}, differently from what happens in the
CDM case, the choice of $\Omega_0$ reflects also in a significant
variation of $\lambda_{FS}$. Therefore, the value of $\Omega_0$ does
not only affect the power spectrum of fluctuations through the
determination of the equivalence epoch, but also changes the
small--scale fluctuation power due to the change in $\lambda_{FS}$.

We plot in Figure \ref{fig:tk} the transfer functions $T(k)$ for density 
perturbations at the outset of recombination for WDM and CDM models,
in the case of $\Omega_0=1$ and $\Omega_0=0.3$ (see caption). This
plot is based on the fitting expression for $T(k)$ provided by
Bardeen et al. \cite{bardeen}. We take $\lambda_{FS}=0.5 \Omega_0^{-1} 
h^{-2}$ Mpc, which is appropriate for light gravitinos
(cf. eq.\ref{eq:free}). As expected, cold and warm scenarios
coincide on scales above $\lambda_{FS}$, while on smaller scales the
fluctuation power is exponentially suppressed for WDM by free--streaming.

As for the scenario {\it (c)}, one can envisage for instance the
possibility that in addition to the warm light gravitinos there are
massive light neutrinos (in the eV range) considering a scheme with a
mixed WDM+ HDM\cite{malaney}. Another possibility, that we have recently
investigated together with M. Yamaguchi \cite{borganietal}, takes into
account the fact that we have a ``secondary population" of gravitinos,
which result from the decay of the next-to-the-lightest superparticle
(NSP), presumably the lightest neutralino. They have a non--thermal
phase--space distribution and exhibit features for the structure
formation which are similar to those of a standard hot light neutrino
in the tens of eV range.  From our analysis it turned out that viable
MDM realizations within the frame of light gravitinos lead to
characteristic features both in the cosmological and particle physics
contexts, making these models testable against astrophysical
observations and future accelerator experiments. In particular, on the
astrophysical side, we find a relatively large $^4$He abundance
(corresponding to slightly more than three neutrino species at
nucleosynthesis), a suppression of high redshift galaxy formation with
respect to the cold dark matter (CDM) scenario and a free-streaming
scale of the non-thermal (``secondary") gravitinos independent of
$m_{3/2}$, but sensitive to the NSP mass (with important consequences
on the large scale structure formation).  As for the particle physics
implications, the implementation of a MDM scheme imposes severe
constraints on the SUSY particle spectrum.  For instance the lightest
neutralino (NSP) should be an essentially pure gaugino and sfermions
have to be rather heavy (in the TeV range).

As a general comment, it is worth stressing once more that, from the
point of view of the formation of cosmic structures, all the scenarios
based on a dominant WDM component differ from their CDM counterpart
only on scales below $\lambda_{FS}$ of gravitinos. Therefore, the
suppression of fluctuation power on galactic scales leads in general
to a later galaxy formation. This is potentially a critical test for
any WDM--dominated scheme, the abundance of high--redshift galaxies
having been already recognized as a non trivial constraints for
several DM models.  It is however clear that quantitative conclusions
on this point would at least require the explicit computation of the
fluctuation power spectrum for the whole class of WDM scenarios, a
project on which we are currently working in collaboration with
E. Pierpaoli.

\vspace{0.3truecm}
\noindent 
{\bf Acknowledgements.} We are grateful to our collaborators,
E. Pierpaoli and M. Yamaguchi, for the work done together on the light
gravitino issue. We also acknowledge stimulating discussions with
S.A. Bonometto and J.R. Primack. We thank the organizers of this
Meeting for the interesting environment in which it took place.  This
work was partially supported by the ``Beyond the Standard Model"
Network, under the EEC Contract No. ERBFMRX-CT96-0090.


\end{document}